\documentstyle[12pt,psfig,bbbold,fullpage]{article}

\def\slantfrac#1#2{\hbox{$\,^{#1}\!/_{#2}$}}

\begin{document}
\thispagestyle{empty}

\hfill{\sf Hot Points in Astrophysics}

\hfill{\sf JINR, Dubna, Russia, August 22-26, 2000}

\vspace*{2.cm}

\begin{center}

{\Large \bf Synthetic Doppler maps of gaseous flows in\\[3mm]
semidetached binaries based on the results\\[5mm] of 3D gas
dynamical simulations}

\end{center}

\vspace*{0.2cm}

\begin{center}

D.V.Bisikalo$^{\dagger}$, A.A.Boyarchuk$^{\dagger}$,
O.A.Kuznetsov$^{\ddagger}$, and V.M.Chechetkin$^{\ddagger}$\\[0.3cm]

{\it $^\dagger$ Institute of Astronomy of the Russian Acad. of Sci., Moscow\\
{\sf bisikalo@inasan.rssi.ru; aboyar@inasan.rssi.ru}\\
$^\ddagger$ Keldysh Institute of Applied Mathematics, Moscow}\\
{\sf kuznecov@spp.keldysh.ru; chech@int.keldysh.ru}

\end{center}

\vspace*{0.2cm}

\begin{abstract}

We present synthetic Doppler maps of gaseous flows in
semidetached binaries based on the results of 3D gas dynamical
simulations. Using of gas dynamical calculations alongside with
Doppler tomography technique permits to identify main features
of the flow on the Doppler maps without solution of ill-posed
inverse problem. Comparison of synthetic tomograms with
observations makes possible both to refine the gas dynamical
model and to interpret the observational data.

\end{abstract}

\vspace*{0.5cm}

\section*{Problem Setup}

Traditional observations of binary systems are carried out using
photometric and spectroscopic methodics. The former gives the
time dependence of brightness in a specific band $I(t)$ and the
latter can give the time dependence of wavelength of some
Doppler-shifted line $\lambda(t)$. Given ephemeris $\varphi(t)$
is known, the dependencies $I(t)$ and $\lambda(t)$ can be
converted with the help of Doppler formula to the light curve
$I(\varphi)$ and phase dependency of radial velocity
$V_R(\varphi)$.

During last ten years the observations of binary systems in the
form of trailed spectrograms for some emission line
$I(\lambda,t)$ or in other terms $I(V_R,\varphi)$ become widely
used. A method of Doppler tomography \cite{MarshHorne88} is
suited to analyze the trailed spectrograms. This method
provides obtaining a map of luminosity in the 2D velocity space
from the orbital variability of emission lines intensity.  The
Doppler tomogram is constructed as a conversion of time
resolved (i.e. phase-folded) line profiles into a map on ($V_x$,
$V_y$) plane. To convert the distribution $I(V_R,\varphi)$ to
Doppler map $I(V_x,V_y)$ we should use the expression for radial
velocity as a projection of velocity vector on the line of
sight, i.e.  $V_R=-V_x\cos(2\pi\varphi)+V_y\sin(2\pi\varphi)$
(here we assume that $V_z\sim0$, the minus sign before $V_x$ is
for consistency with the coordinate system), and solve an
inverse problem that is described by integral equation (see
Appendix A of \cite{MarshHorne88}):

\[
I(V_R,\varphi)=\int\int
I(V_x,V_y)g(V_R+V_x\cos(2\pi\varphi)-V_y\sin(2\pi\varphi))
dV_x dV_y\,,
\]
where $g(V)$ is normalized local line profile shape (e.g., a
Dirac $\delta$-function), and the limits of integration are from
$-\infty$ to $+\infty$. This inverse problem is ill-posed and
special regularization is necessary for its solving (e.g.
Maximum Entropy Method \cite{MarshHorne88,NarayanNit86}, Fourier
Filtered Back Projection \cite{Robinson93}, Fast Maximum Entropy
Method \cite{Spruit98}, etc., see also \cite{Frieden79}). As a
result we obtain a map of distribution of specific line
intensity in velocity space.  This map is easier to interpret
than original line profiles, moreover, the tomogram can show
(or at least gives a hint to) some features of flow structure.
In particular, the double-peaked line profiles corresponding to
circular motion of the gas (e.g. in accretion disk) become a
diffuse ring-shaped region in this map.  Resuming, we can say
that components of binary system can be resolved in velocity
space while they can not be spatially resolved through direct
observations, so the Doppler tomography technique is a rather
power tool for studying of binary systems.

Unfortunately the reconstruction of spatial distribution of
intensity on the basis of Doppler map is an ambiguous problem
since points located far from each other may have equal radial
velocities and deposit to the same pixel on the Doppler map. So
the transformation $I(V_x,V_y)\to I(x,y)$ is impossible without
some {\it a priori} assumptions on the velocity field.

The situation changes drastically when one uses gas dynamical
calculations alongside with Doppler tomography technique. In
this case we are not need to cope with the inverse problem since
the task is solved directly: $\rho(x,y)~\&~T(x,y)\to I(x,y)$ and
$I(x,y)~\&~V_x(x,y)~\&~V_y(x,y)\to I(V_x,V_y) \to
I(V_R,\varphi)$.  Difficulties can arise when converting the
spatial distributions of density and temperature $\rho(x,y)$,
$T(x,y)$ into the distribution of luminosity of specific
emission line $I(x,y)$. For thick lines the formation of line
profile should be described by radiation transfer equations
(see, e.g., \cite{HorneMarsh86}), therefore to make our
preliminary synthetic Doppler maps we assume that the matter is
optically thin. As a first approximation we adopt the line
luminosity as $I=\rho$ and as $I=\rho^2 T^{1/2}$
\cite{Richards98,Ferland80}.

We also should emphasize the complexity of Doppler maps
analysis for eclipsing binaries (see, e.g., \cite{Kaitchuck94}).
The synthetic Doppler map produces the line emissions from any
site of binary and suggests that they are visible at all orbital
phases, in other words there are no eclipses and occultations of
emission regions. Clearly, eclipsing systems violates that
assumption. Usually, when dealing with observations, the
eclipsed parts of trailed spectrograms are naturally excluded
from input data for construction of Doppler tomograms. But
conversion of gas dynamical simulation results into Doppler maps
suggests using of {\it full} set of data. Thus when analyzing of
synthetic Doppler maps for eclipsing binaries we should take in
mind that some features can be occulted on some phases even for
optically thin case.

\begin{figure}[t]
\centerline{\hbox{\psfig{figure=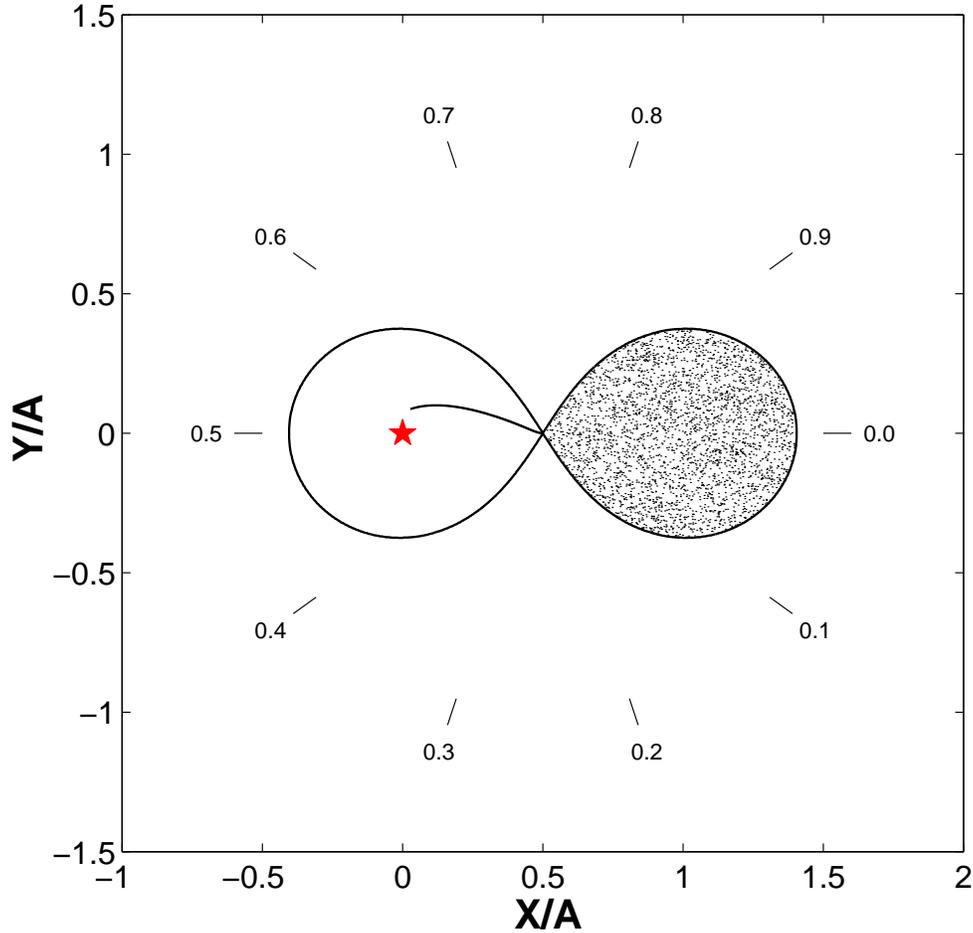,width=13cm}}}
\caption{\footnotesize The adopted coordinate system with phase
angles of observer in a binary system. The $x$ and $y$
coordinates are expressed in terms of the separation $A$. The
red asterisk is the accretor. The donor-star is shadowed.
The critical Roche lobe and ballistic trajectory of a particle
moving from $L_1$ are shown by a solid lines. Orbital rotation
of the binary is counter-clockwise.} \end{figure}

\begin{figure}[t]
\centerline{\hbox{\psfig{figure=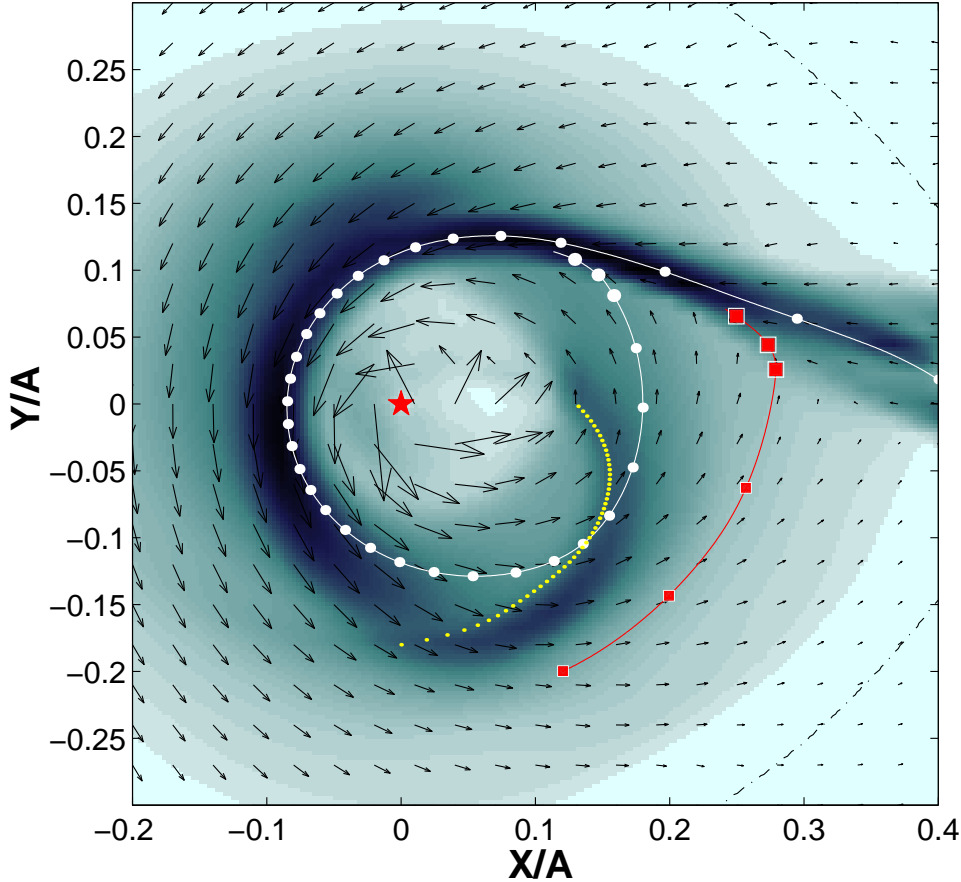,width=13cm}}}
\caption{\footnotesize The distribution of density over the
equatorial plane. The $x$ and $y$ coordinates are expressed in
terms of the separation $A$. Arrows are the velocity vectors in
observer's frame. The red asterisk is the accretor. The
dashed-dotted line is Roche equipotential passing through $L_1$.
The yellow dotted line is the tidally induced spiral shock. Gas
dynamical trajectory of a particle moving from $L_1$ to accretor
is shown by a white line with circles. Another gas dynamical
trajectory is shown by a red line with squares (see also
Fig.~6).} \end{figure}

\begin{figure}[t]
\centerline{\hbox{\psfig{figure=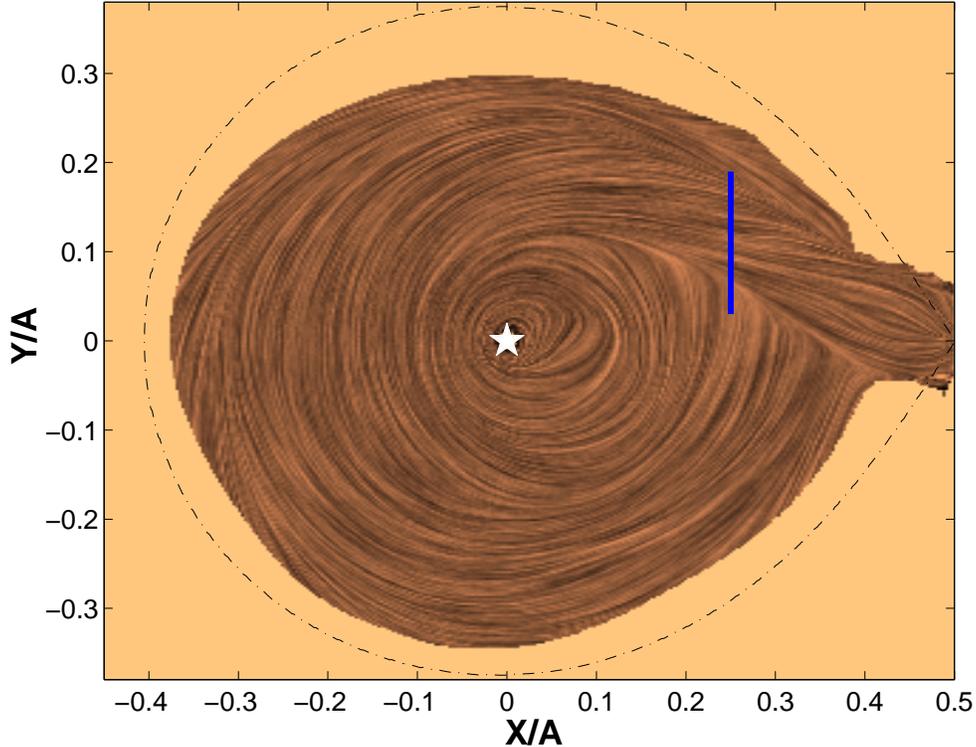,width=13cm}}}
\caption{\footnotesize The texture of flow the equatorial plane
(in corotation frame), i.e. visualization of velocity vectors
field using Line Integral Convolution Method \protect\cite{LIC}.
The white asterisk is an accretor. See Fig.~4 for the
explanation of a blue line.} \end{figure}

\begin{figure}[t]
\centerline{\hbox{\psfig{figure=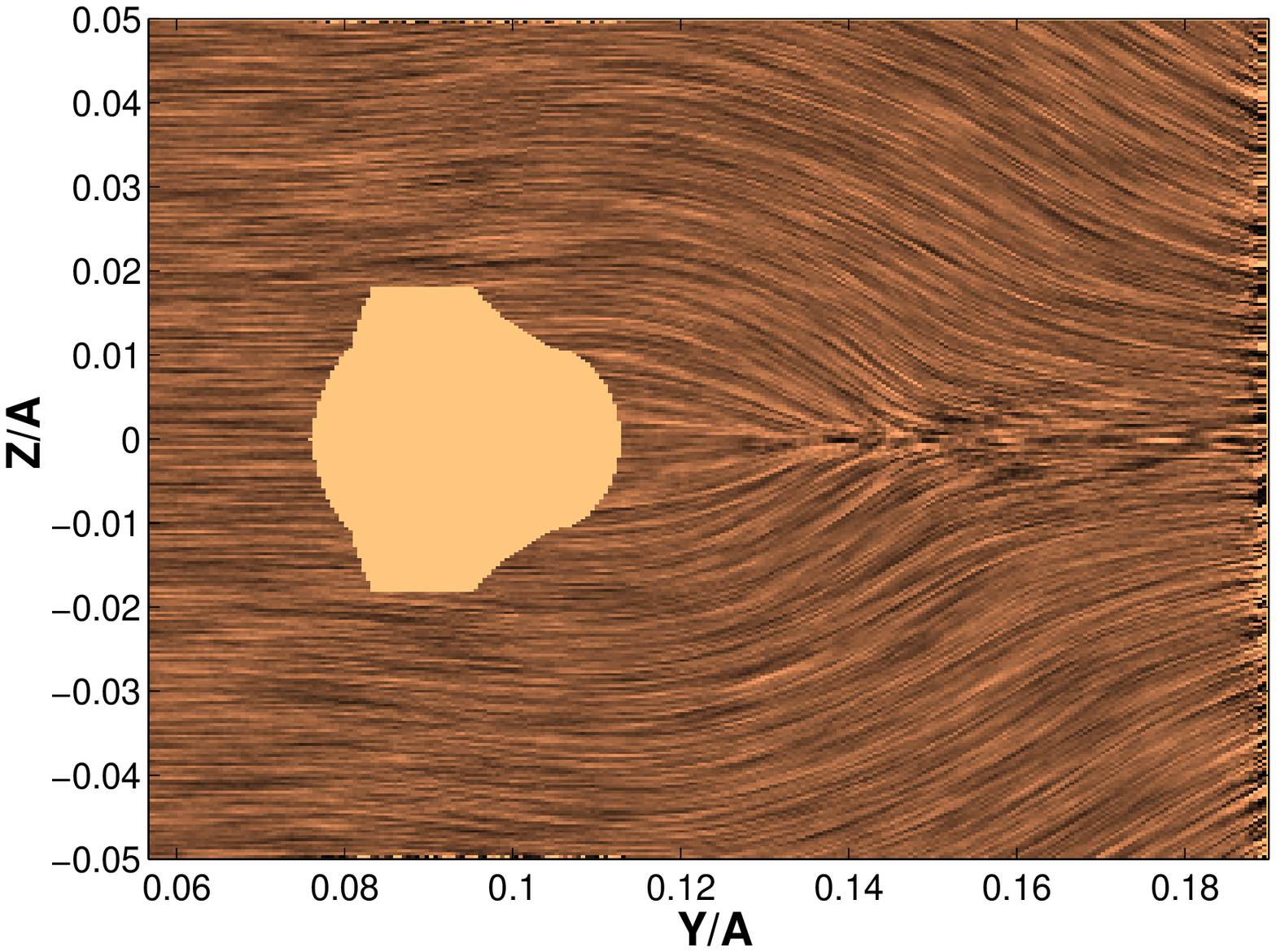,width=13cm}}}
\caption{\footnotesize The texture of flow the $YZ$ slice along
a blue line in Fig.~3. Empty region corresponds to
cross-section of the stream from $L_1$.} \end{figure}

\section*{3D gas dynamical simulations: the model and results}

The full description of the used 3D gas dynamical model can be
found in \cite{lowvisc}. Here we pay attention only to the main
features of the model.

Let us consider the semidetached binary system with mass of
accretor $M_1$, mass of donor-star $M_2$, separation $A$, and
velocity of orbital rotation $\Omega$. These parameters are
connected by third Kepler's low $A^3\Omega^2=G(M_1+M_2)$. To
describe the gas flow in this binary system we used the 3D
system of Euler equations. The calculations were carried out in
the non-inertial Cartesian coordinate system rotating with the
binary system. To close the system of equations, we used the
equation of state of ideal gas $P=(\gamma-1)\rho\varepsilon$,
where $\gamma$ is the ratio of heat capacities. To mimic the
system with radiative losses, we accept in the model the value
of adiabatic index close to unit:  $\gamma = 1.01$, that
corresponds to the case close to the isothermal one
\cite{Sawada86,isoterm}.

To obtain numerical solution of the system of equations we used
the Roe--Osher TVD scheme of a high approximation order
\cite{Roe,Osher} with Einfeldt modification \cite{Einfeldt}. The
original system of equations was written in a dimensionless
form. To do this, the spatial variables were normalized to the
distance between the components $A$, the time variables were
normalized to the reciprocal angular velocity of the system
${\Omega}^{-1}$, and the density was normalized to its
value\footnote{Because the system of equations can be scaled to
density and pressure, the density scale was chosen simply for
the sake of convenience.} in the inner Lagrangian point $L_1$.
We adopted the computational domain as a parallepipedon
$[-\slantfrac{1}{2}A\ldots
\slantfrac{1}{2}A]\times[-\slantfrac{1}{2}A\ldots
\slantfrac{1}{2}A]\times[0\ldots \slantfrac{1}{4}A]$
(due to the symmetry of the problem calculations were conducted
only in the top half-space). A sphere with a radius of
$\slantfrac{1}{100}A$ representing the accretor was cut out of
the calculation domain. The boundary conditions were taken as
`free outflow' on the accretor star and on the outer edges of
computational domain. In gridpoint corresponding to $L_1$ we
injected the matter with parameters $\rho=\rho(L_1)$,
$V_x=c(L_1)$, $V_y=V_z=0$, where $c(L_1)$ is a gas speed of
sound in $L_1$ point. As the initial conditions we used rarefied
gas with the following parameters
$\rho_0=10^{-5}\cdot\rho(L_1)$,
$P_0=10^{-4}\rho(L_1)c^2(L_1)/\gamma$, ${\bmath V}_0=0$.

Analysis of considered problem shows that the gas dynamical
solution is defined by three dimensionless parameters
[16--18]: mass ratio $q=M_2/M_1$, Lubow-Shu
parameter $\epsilon=c(L_1)/A\Omega$ \cite{LubowShu75}, and
adiabatic index $\gamma$. The value of adiabatic index was
discussed above and we used the value $\gamma=1.01$. Analysis of
our previous results \cite{param,2D3D} shows that the main
characteristic features of 3D gas dynamical flow structure are
qualitatively the same in wide range of parameters $q$ and
$\epsilon$.  Therefore for model simulation we chose them as
follows:  $q=1$, $\epsilon=\slantfrac{1}{10}$.


The results of calculations will be presented in coordinate
system (see Fig.~1) which is widely used in Doppler mapping.
The origin of coordinates is located in the center of the
accretor, `$x$'-axis is directed along the line connecting the
centers of stars, from accretor to the mass-losing component,
`$z$'-axis is directed along the axis of rotation, and
`$y$'-axis is determined so that we obtain a right-hand
coordinate system (i.e. `$y$'-axis points in the direction of
orbital movement of the donor-star). In Fig.~1 we put digits
showing the phase angles of the observer in binary system. We
also show a critical Roche lobe with shadowed donor-star and
ballistic trajectory of a particle moving from $L_1$ point to
accretor.

The morphology of gaseous flows in considered binary system can
be evaluated from Figs~2, 3 and 4. In Fig.~2 the distribution of
density over the equatorial plane and velocity vectors are
presented. In this Figure we also put a gas dynamical trajectory
of a particle moving from $L_1$ to accretor (a white line
with circles) and a gas dynamical trajectory passing through the
shock wave along the stream edge (a red line with
squares, see also Fig.~6). In Fig.~3 we present the so called
texture figure in equatorial plane which is visualization of
velocity vectors field using Line Integral Convolution Method
\cite{LIC}. In Fig.~4 the similar texture is presented for $YZ$
slice along a blue line in Fig.~3. Empty region in Fig.~4
corresponds to cross-section of the stream.

Analysis of presented results as well as our previous studies
\cite{paper1,mnras} show the significant influence of rarefied
gas of circumbinary envelope on the flow patterns in
semidetached binaries. The gas of circumbinary envelope
interacts with the stream of matter and deflects it. This leads,
in particular, to the shock-free (tangential) interaction
between the stream and the outer edge of forming accretion disc,
and, as the consequence, to the absence of `hot spot' in the
disc. At the same time it is seen, that the interaction of the
gas of circumbinary envelope with the stream results in the
formation of an extended shock wave located along the stream
edge (`hot line'). From Figs~2 and 3 it also seen that spiral
shock tidally induced by donor-star appears (dotted line in
Fig.~2). Appearance of tidally induced two-armed spiral shock
was discovered by Matsuda \cite{Sawada87,spiral1,spiral2}. Here
we see only one-armed spiral shock. In the place where the
second arm should be we see the strong gas dynamical interaction
of circumbinary envelope with the stream causing the formation
of rather intensive shock along the outer edge of the stream.
This gas dynamically induced shock probably prevents the
formation of second arm of tidally induced spiral shock.

An analysis of the flow structure in $YZ$ plane (Fig.~4) shows
that a part of the circumstellar envelope interacts with the
(denser) original gas stream and is deflected away from the
orbital plane. This naturally leads to the formation of `halo'.
Following to \cite{lowvisc}, one can define `halo' as that
matter which: i) encircles the accretor being gravitationally
captured; ii) does not belong to the accretion disc; iii)
interacts with the stream (collides with it and/or overflows
it); iv) after the interaction either becomes a part of the
accretion disc or leaves the system.

Described gas dynamical features of flow structure are good
candidates to be observed using Doppler mapping technique.

\begin{figure}[t]
\centerline{\hbox{\psfig{figure=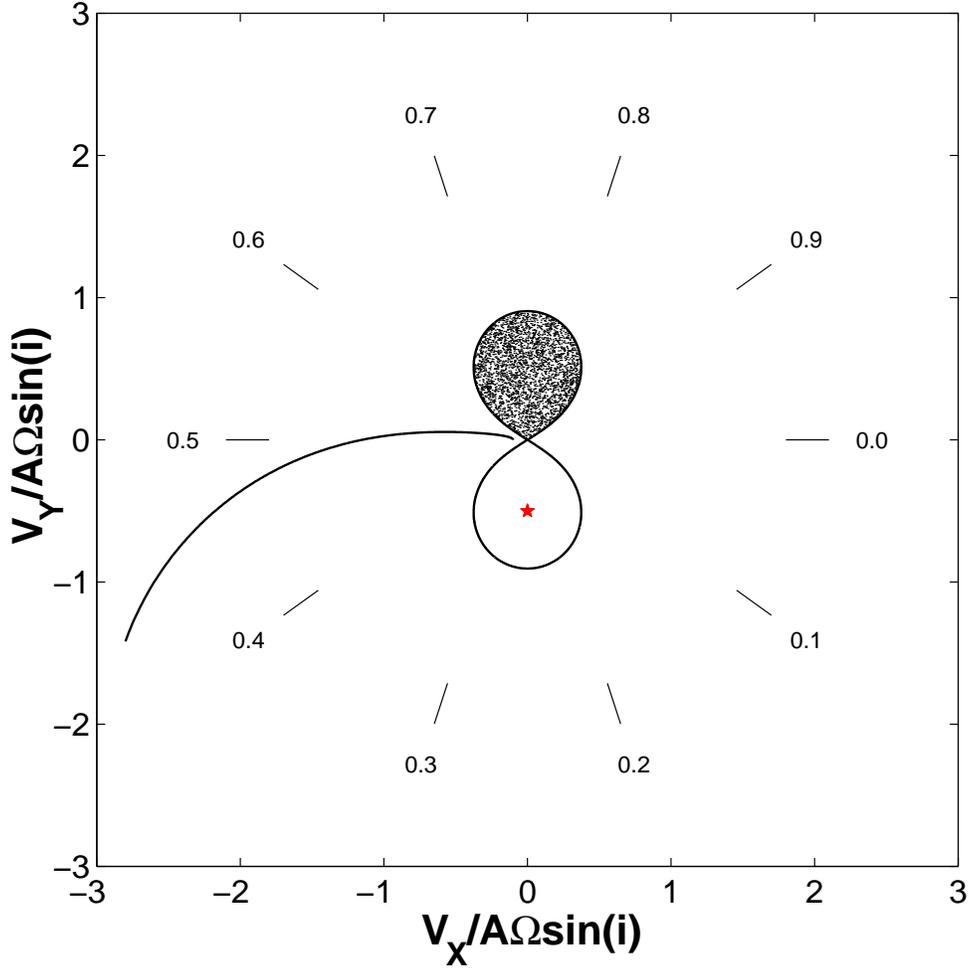,width=13cm}}}
\caption{\footnotesize The adopted velocity coordinate system
with phase angles of observer in a binary system. The critical
Roche lobe and ballistic trajectory of a particle moving from
$L_1$ point to accretor are shown by a solid line.
The red asterisk is the accretor. The donor-star is
shadowed.} \end{figure}

\section*{Synthetic Doppler maps: methodics}

The Doppler maps show the distribution of luminosity in
the velocity space. Each point of flow has a three-dimensional
vector of velocity ${\bmath U}=(U_x,U_y,U_z)$ in observer's
(inertial) frame. In the case when observer is located in the
orbital plane of the binary the Doppler map's coordinate
$(V_x,V_y)$ will coincide with $U_x$ and $U_y$. To define these
coordinates for the case of inclined system we have to find a
projection of vector ${\bmath U}$ on the plane constituted by
vectors $\bmath n$ and ${\bmath n}\times{\bmath\Omega}$, where
$\bmath n$ is a direction from the observer to binary.

The line emissivity in the velocity space can be written as:

\[
I(V_x,V_y)\sim\int\limits_{\cal
O}\int\limits_{U_x}\int\limits_{U_y} I(x,y,z)
\delta(U_x(x,y,z)\sin i+U_z(x,y,z)\cos i-V_x)
\]
\[\qquad\qquad
\qquad\qquad
\delta(U_y(x,y,z)\sin i+U_z(x,y,z)\cos i-V_y)
d{\cal O}dU_xdU_y\,,
\]
where $d{\cal O}=dxdydz$, $i$ -- inclination angle. Usually
$U_z\ll U_x,U_y$ in the most dense parts of the flow therefore
we can neglect the third component of velocity $U_z$. This
assumption allows us to take off `$\sin i$' beyond the integral
and to simplify the expression. Using $V_x/\sin i$ and $V_y/\sin
i$ as coordinates for Doppler map will hide the dependency on
the inclination angle.

The adopted dimensionless coordinate system for Doppler maps is
shown in Fig.~5. We also put in Fig.~5 digits (the same as in
Fig.~1) showing the phase angles of observer in binary system.
The transformation of the donor-star from spatial to velocity
coordinate system is very simple as it is fixed in the
corotating frame. Every point $\bmath r$ fixed in the binary
frame has a velocity ${\bmath\Omega}\times{\bmath r}$ in
corotation frame. This is linear in the perpendicular distance
from the rotation axis and therefore the shape of the donor-star
projected on the orbital plane is preserved (see Fig.~5). Since
the velocity of each point of donor-star is perpendicular to the
radius vector, all points of donor-star are rotated by
90$^\circ$ counter-clockwise between the spatial (see Fig.~1)
and velocity coordinate diagrams \cite{MarshHorne88}.

Figure~5 depicts in velocity coordinates a critical Roche lobe
with shadowed donor-star and ballistic trajectory of a particle
moving from $L_1$ to accretor. On the velocity plane the
accretor has coordinates (0,$K_1$), where $K_1=-A\Omega
M_2/(M_1+M_2)$ or in dimensionless form (for adopted value of
$q=1$) $K_1=-\slantfrac{1}{2}$.

\begin{figure}[t]
\centerline{\hbox{\psfig{figure=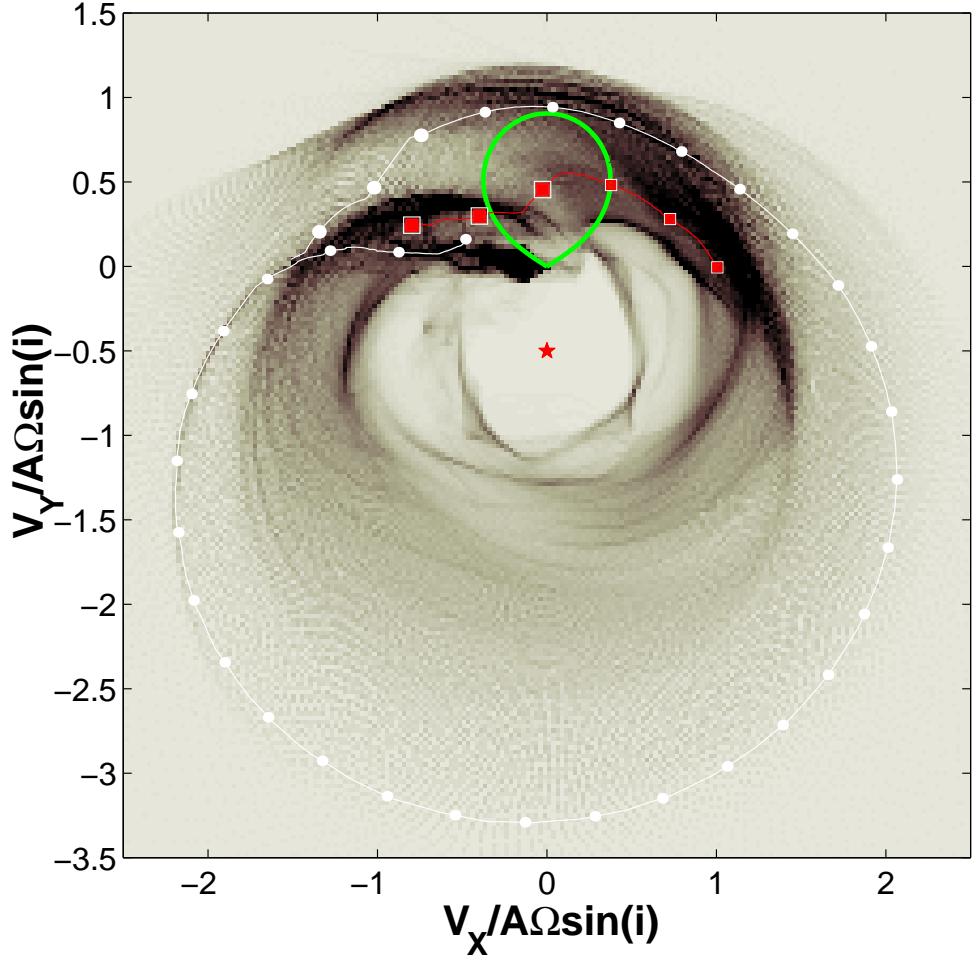,width=13cm}}}
\caption{\footnotesize Synthetic Doppler map for
$I=\rho$. The secondary Roche lobe (a bold green line) and
the accretor (a red asterisk) are also shown. The white
line with circles and red line with squares show gas dynamical
trajectories in the velocity coordinates (see Fig.~2).}
\end{figure}

\begin{figure}[t]
\centerline{\hbox{\psfig{figure=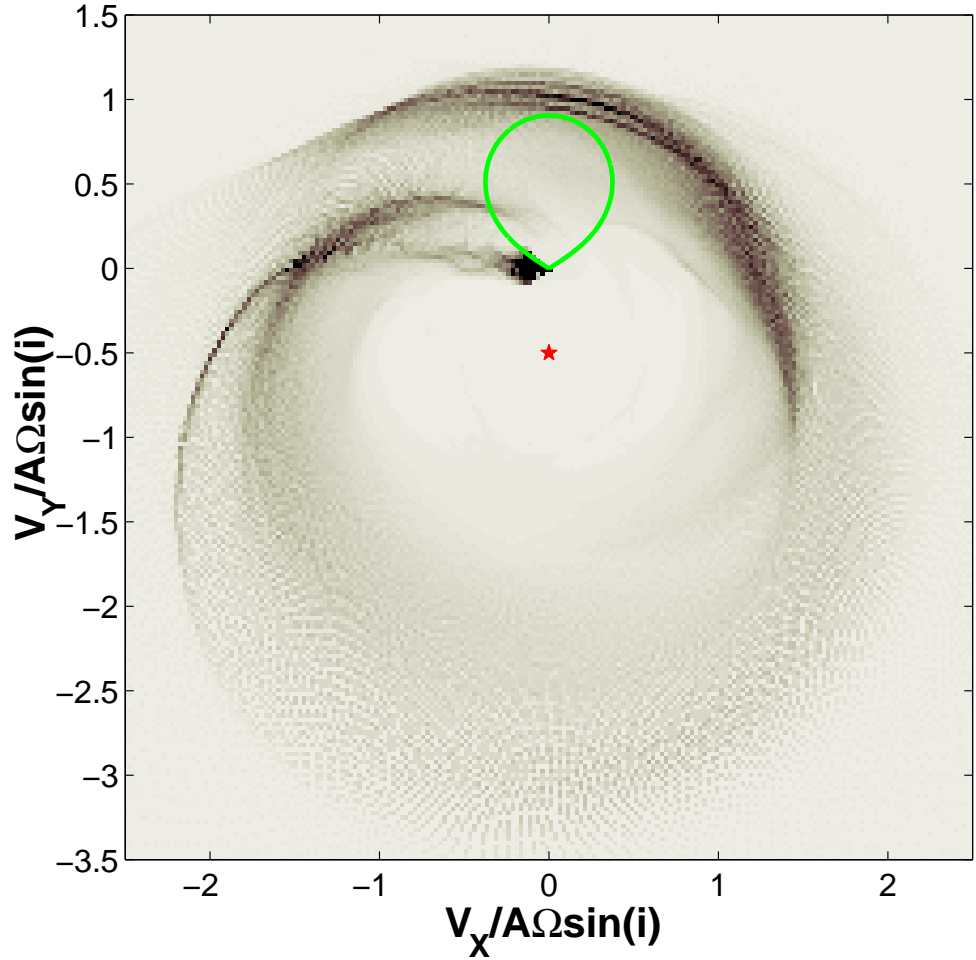,width=13cm}}}
\caption{\footnotesize Synthetic Doppler map for
$I=\rho^2T^{1/2}$.} \end{figure}

\begin{figure}[p]
\centerline{\hbox{\psfig{figure=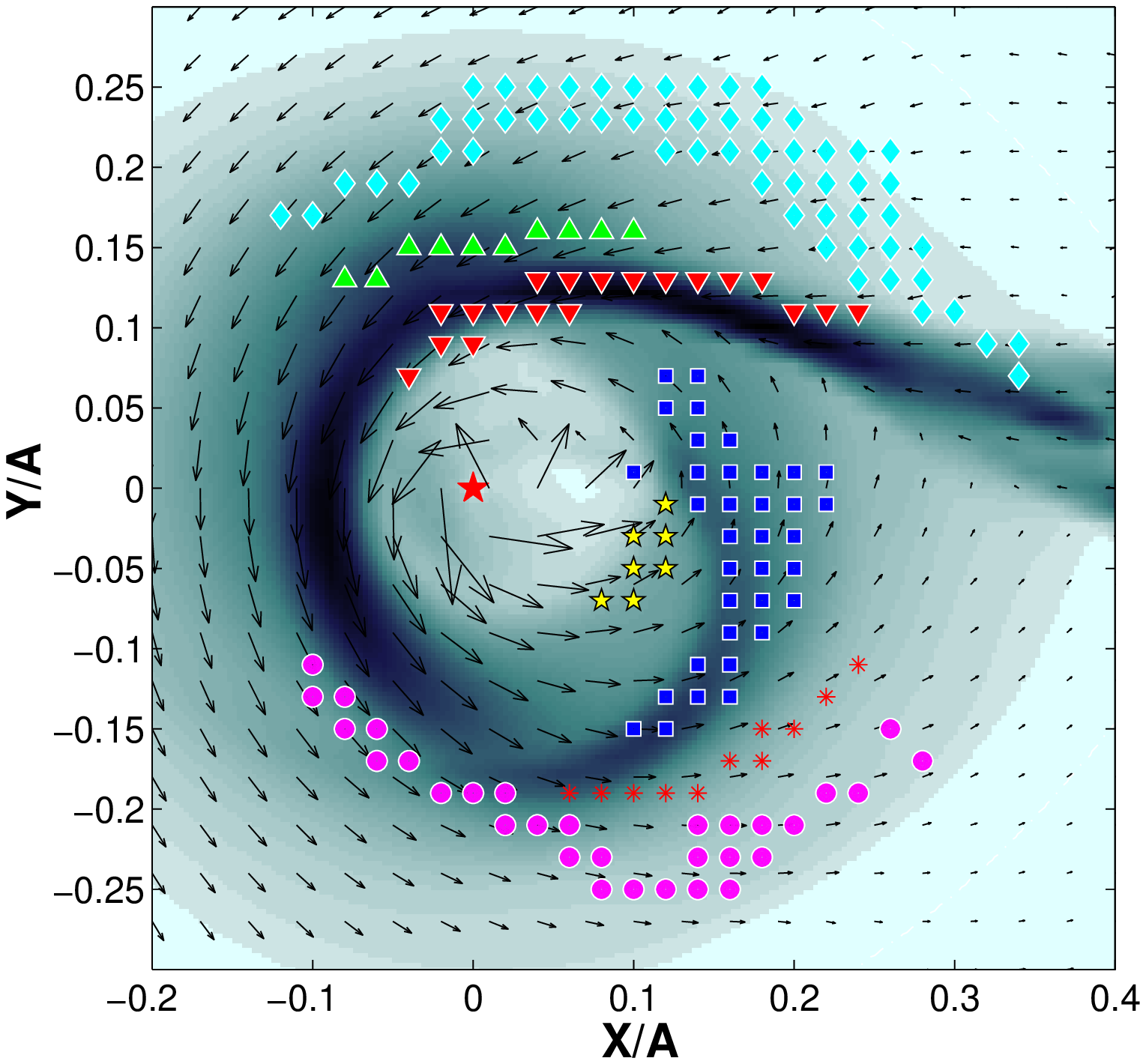,width=12cm}}}
\vspace{3mm}
\centerline{\hbox{\psfig{figure=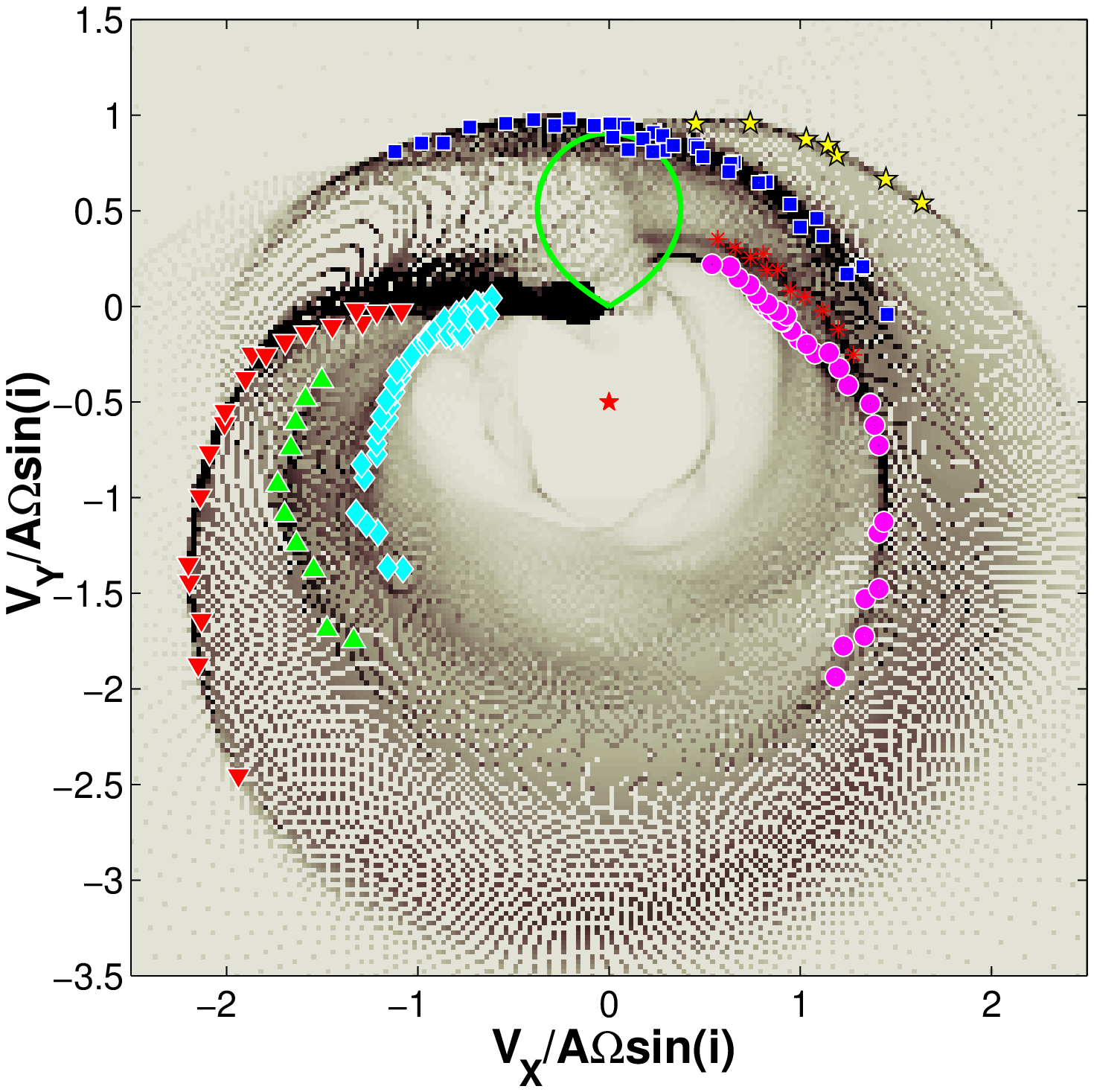,width=12cm}}}
\caption{\footnotesize (Upper panel) Density and
velocity distribution in equatorial plane with marked sites.
(Lower panel) Synthetic Doppler map corresponding to the equatorial
plane with the same marks.} \end{figure}

\section*{Synthetic Doppler maps: analysis}

As it was mentioned earlier we used both $I=\rho$ and
$I=\rho^2T^{1/2}$ to build synthetic Doppler maps.  The
synthetic Doppler map calculated with $I=\rho$ is presented in
Fig.~6. This map shows the intensity integrated over the
$z$-coordinate, i.e. taking into account all $z$-layers of
computational grid. Figure~7 presents the Doppler map
corresponding to $I=\rho^2T^{1/2}$. The comparison of results
presented in Figs~6 and 7 shows that all characteristic features
are the same for different prescriptions of $I$, therefore for
further analysis we will use results for $I=\rho$.

For the sake of comparison we have marked different features of
flow structure and put these marks both in spatial and on
velocity coordinates. Let us consider results presented in
Figs~2 and 6. As it follows from these figures stream from $L_1$
(first part of curve marked by white line with circles)
transforms to a spiral arm in the 2-nd and 3-d quadrants of the
Doppler map. Shock wave caused by interaction of circumbinary
envelope with the stream is located along the stream edge in
spatial coordinates. Three last points (marked by larger
symbols) of curves with circles and squares in Fig.~2 are the
examples of two flowlines passing through the shock. The
position of this shock on Doppler map is an spiral arm starting
approximately from the center of donor-star in the direction of
negative $V_x$. This arm lies above the arm caused by stream
from $L_1$.

To analyze other features we have marked the corresponding sites
both in spatial and velocity coordinates (see Fig.~8a,b). This
comparison is a straightforward problem since we simply
choose from all sites of gas dynamical flow structure those that
fall in specific places on the Doppler map and mark them. It is
seen that tidally induced spiral shock (dashed line in Fig.~2)
or more precisely the dense post-shock zone (blue squares in
Fig.~8a) produces bright arm in the first quadrant of Doppler
map. It is important to note that gas of circumbinary envelope
overflowing the stream (cyan diamonds and green up triangles)
significantly increase the luminosity of the zone of Doppler map
located below the zone corresponding to the stream. Analysis of
other features can be conducted by comparison of Figs~8a and 8b
as well (magenta circles, yellow stars and red asterisks).

\section*{Doppler maps: observations}

Doppler tomography of binaries is widely and actively used now.
Since pioneer work by Marsh and Horne \cite{MarshHorne88} a lot
of Doppler map observations were made (see, e.g., [25--28]).
Just to give to reader the idea on observable
Doppler maps we present here in Fig.~9 two Doppler maps of
{\sc HeII}~$\lambda~4686$\AA~ for IP Peg obtained by
Morales-Rueda, Marsh and Billington \cite{Luisa}.

\begin{figure}[t]
\centerline{\hbox{\psfig{figure=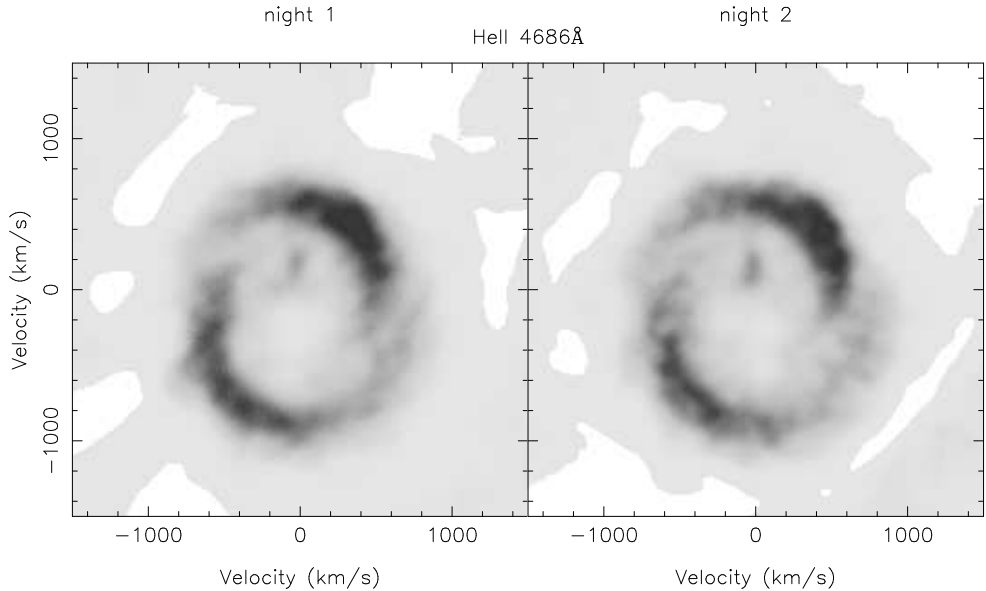,angle=-90,width=13cm}}}
\caption{\footnotesize Doppler maps of
{\sc HeII}~$\lambda~4686$\AA~ for IP Peg \protect\cite{Luisa}.
This figure is reproduced under the kind permission by
L.Morales-Rueda.} \end{figure}

\section*{Conclusions}

Three-dimensional gas dynamical simulations of flow structure in
semidetached binaries allow to build  synthetic Doppler maps. In
turn it gives a possibility to identify main features of the
flow on the Doppler map without solution of ill-posed inverse
problem. Comparison of synthetic tomograms with observations
makes possible both to refine the gas dynamical model and to
interpret the observational data.

\section*{Acknowledgments}
This work was supported by the Russian Foundation for Basic
Research (grants 99-02-17619, 00-01-00392, 00-02-17253) and by
grants of President of Russia (99-15-96022, 00-15-96722).

\end{document}